\documentclass[aps,prd,reprint,showpacs,nofootinbib]{revtex4-1}

\usepackage[english]{babel}
\usepackage{graphicx}
\usepackage{array}
\usepackage{dcolumn}
\usepackage{bm}
\usepackage{amssymb}
\usepackage{amsmath}
\usepackage{amsfonts}
\usepackage{amsbsy}

\def\scr{$\tilde{\tau}_{\text{CR}}$}
\def\fluorine{$^{19}$F}
\def\iodine{$^{127}$I}
\def\germanium{$^{73}$Ge}

\begin{document}

\title{
On the formalism and upper limits for spin-dependent cross sections
in dark matter elastic scattering with nuclei }

\author{M.~Cannoni}
\email{Email:mirco.cannoni@dfa.uhu.es}
\affiliation{Departamento de F\'isica Aplicada, Facultad de Ciencias
Experimentales, Universidad de Huelva, 21071 Huelva, Spain}

\begin{abstract}

We revise the spin-dependent neutralino-nucleus elastic scattering comparing the formalisms 
and approximations found in literature for the momentum transfer dependent 
structure functions.
We argue that one of the normalized structure functions of Divari, Kosmas, Vergados and Skouras 
is all that one needs to correctly take into account 
the detailed nuclear physics information provided by shell-model calculations.
The factorization of the particle physics 
degrees of freedom from the nuclear physics momentum dependent structure functions
implied by this formalism allows for a better understanding of the so-called
model independent method for setting upper limits.
We further discuss the possibility of experiments with spin-dependent sensitivity like COUPP 
to test or set limits on the proton spin-dependent cross section in the framework 
of the stau co-annihilation region of the constrained minimal supersymmetric standard
model. 
For this  model with $A_0 =0$, we provide a fitting formula by which it is possible 
to convert an upper limit on the spin-independent cross section as a function of the neutralino mass 
directly into an exclusion plot in the ($m_{1/2}$, $\tan\beta$) plane.

\end{abstract}

	
\date{October 17, 2011 V2}

\maketitle

\section{Introduction}
\label{sec1}

The nature of non-baryonic dark matter that seems to constitute the largest part
of the matter in the Universe is still unknown. If dark matter is formed by  
non-relativistic weakly interacting massive particles (WIMP) distributed in 
the halo of the galaxy, they should scatter elastically with the nuclei in 
a terrestrial detector~\cite{gw}. 
A characteristic signal of the WIMP interaction is the presence of an annual 
modulation in the event rate correlated with the motion of the Earth~\cite{freese}. 

Experimental evidence of this 
modulation has been reported in the last years by the DAMA 
collaboration~\cite{dama}, and recently, also by the CoGENT collaboration~\cite{cogent}. 
The interpretation of these signals favors a light WIMP with mass around 10 GeV and a 
large spin-independent (SI) WIMP--nucleon cross section of order of $10^{-4}$ pb~\cite{hooper}. 
Other experiments, CDMS~\cite{cdms}, XENON100~\cite{xenon100} and SIMPLE~\cite{simple}, 
that anyway are not sensitive to the annual modulation, have reported upper limits that 
challenge the values of the cross section and mass statistically favored by DAMA and CoGENT.

If on the experimental side the situation is at least
controversial~\cite{collarcritics,simplereply},
on the theoretical side it is not less ambiguous.	 
In the popular scheme of the minimal supersymmetric standard model (MSSM)
with R-parity conservation where the lightest neutralino is a natural WIMP 
candidate, it is possible to accommodate a light neutralino with a cross section able to explain DAMA and 
CoGENT results while not contradicting other phenomenological 
constraints~\cite{fornengo,belli,deaustri,calibbi}.

In supersymmetric models with unification conditions like the constrained MSSM (CMSSM)
light neutralinos with such a large spin-independent cross section 
are excluded by other experimental constraints such as the LEP bound on the chargino mass. 
On the other hand, global fits that take into account accelerator, flavor physics
and dark matter constraints, single out best fit points of the parameter space with a heavy
neutralino~\cite{ellisfit,farina,bertone}.  

In this paper we thus consider a region of the CMSSM parameter space, 
the so-called stau co-annihilation region (\scr). 
In particular, we are interested to find out if present 
and future experiments can constrain this region by the spin-dependent (SD) 
elastic scattering.

In the case of WIMP like the lightest neutralino 
(or any candidate with the same structure of coupling with nucleons), 
setting constrains on the SD couplings is, confronted with the SI case, 
complicated by the fact that: 
$(a)$ there are two elementary cross sections, WIMP-proton and WIMP-neutron,
that in principle should be constrained at the same time and 
in a way that does not depend on the neutralino ``composition'' (the SI proton and neutron 
cross sections are to a very good approximation equal); 
$(b)$ in the formula for the neutralino-nucleus cross section  the particle physics degrees of freedom 
are not factorized from the momentum dependent spin structure functions (SSF), 
thus when setting upper limits one is forced to fix the neutralino ``composition'' by the ratio of the couplings. 
Actually, problem  $(b)$ is at the root of problem $(a)$.

A solution to the problem $(a)$ has been proposed in Ref.~\cite{tovey}.
Thereafter the method has become the standard way to derive limits on the SD 
WIMP-nucleon cross sections and to combine them from different experiments~\cite{giuliani,giuliani1}.

We have discussed problem $(b)$ in a previous paper~\cite{Cannoni} where it is evidenced
that the foreseen factorization is actually achieved by simply normalizing the standard 
structure functions to their value at zero momentum transfer.

Here we show that the solution of problem $(b)$ indeed gives 
a better understanding of the solution to the problem $(a)$
proposed by Ref.~\cite{tovey}. In particular we show that
there is no need of the assumptions
made in Re.~\cite{tovey} that were object of 
criticisms Refs.~\cite{bednyakov_aspects,BK}. The method is 
not limited to the zero momentum transfer cross section but actually 
can incorporate the full momentum dependent structure functions. 
This is done in Section~\ref{sec3}.

In Section~\ref{sec2},  and in the Appendix,
we discuss various aspects of the momentum transfer dependent formalism
and argue that some unnecessary complications of the standard formalism
are at the origin of the aforementioned problems.

In Section~\ref{sec4} we then discuss
to what extent the limits on the single WIMP-nucleon cross sections derived 
by actual experiments like COUPP and XENNON100  can constraint the \scr.

In Section~\ref{sec5} we give a parametrization of the SI
neutralino-nucleon cross section in the stau-co-annihilation region that allows
to translate an experimental upper limit into a bound in the ($m_{1/2}$, $\tan\beta$)
plane.

Summary and conclusions are given in Section~\ref{sec6}.
In the Appendix we provide a detailed derivation of the formulas 
discussed in Section~\ref{sec2}.

\begin{figure*}[t!]
\includegraphics*[scale=0.71]{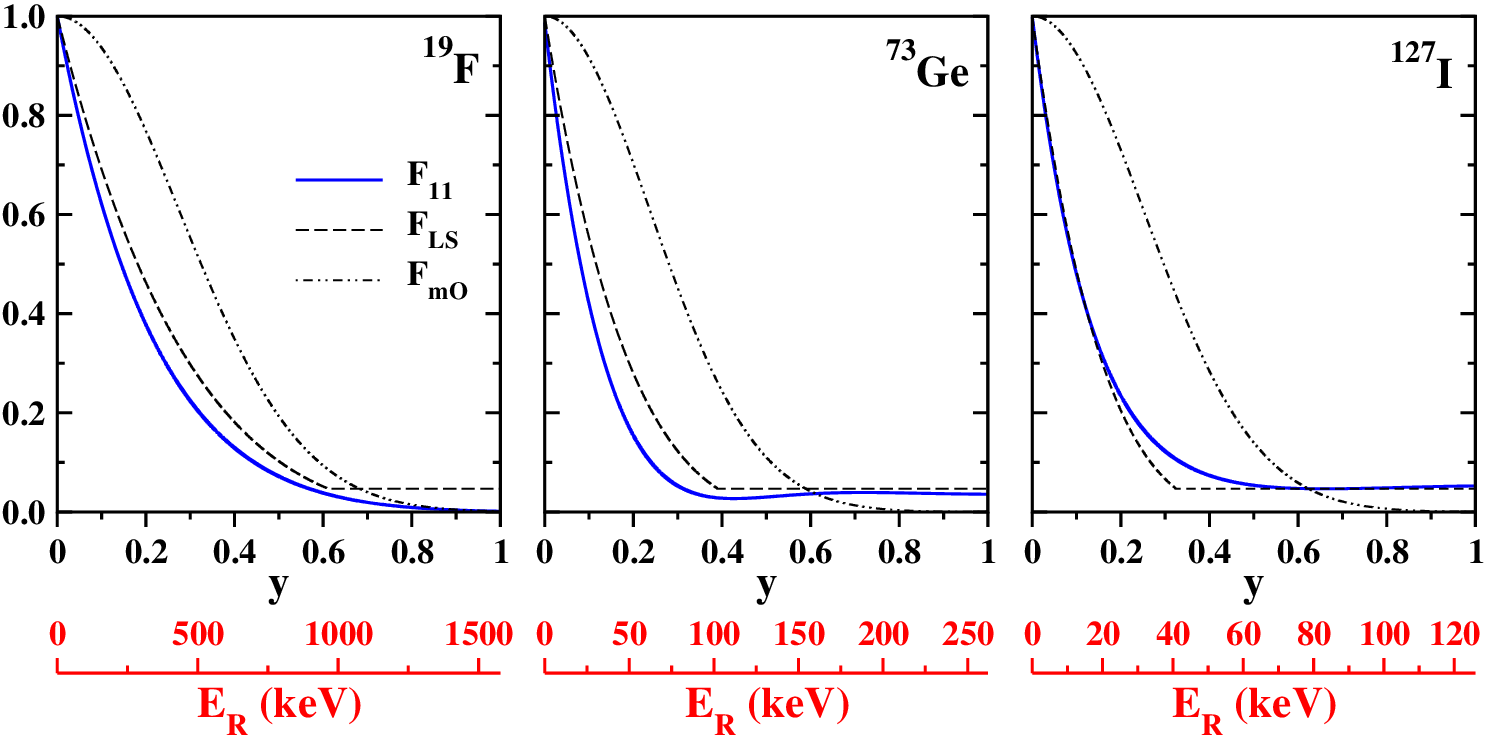}
\caption{In blue line the normalized structure function $F_{11}=S_{11}(q)/S_{11}(0)$, 
for the nuclei \fluorine, \germanium  ~and \iodine.
The dashed line refers to the parametrization of Eq.~(\ref{FSDLS}) and the dashed-dotted line to 
the parametrization of Eq.~(\ref{FSDmicr}). The variable in the abscissas is $y=(qb/2)^2$,
being $q$ the momentum transfer and $b=1\text{fm}A^{1/6}$ the oscillator size parameter. 
In the red abscissas the corresponding values of the recoil energies in keV are given.}
\label{fig1:SSF}
\end{figure*}

\section{SD formalism revised}\label{sec2}
\subsection{Structure functions and ``form factors'' }
\label{sec2.1}

Direct detection experiments employing odd nuclei with non-zero ground state
angular momentum $J$, aim to constrain, in the case of absence of a positive signal, 
the spin-spin interaction of dark matter particles with the nucleons. 
Detailed nuclear shell-model calculations of the spin matrix elements in the
zero momentum transfer limit (ZMTL), i.e. point-like nucleus, 
and of the SSF that account for the momentum transfer 
dependence response, have been carried out for many nuclei employed in 
actual experiments, see~\cite{BS1} for reviews.

The differential neutralino-nucleus cross section,
as function of the recoil energy of the nucleus $E_R=q^2 /2m_A$ being $q$
the modulus of the momentum transfer,
has the general form
\begin{equation}
\frac{d\sigma^{\lambda}_{A}}{dE_R} =\frac{m_A}{2\mu_A^2 v^2}
\sigma^{\lambda}_{A} (0) \Phi^{\lambda} (E_R ).
\label{dsiggen}
\end{equation}
Here $\lambda =\text{SI}$ or $\lambda =\text{SD}$, $m_A$ is the mass of the nucleus with mass number 
$A$, 
$\mu_A$ the neutralino-nucleus reduced mass and $v$ the relative velocity.
$\sigma^{\lambda}_{(A)} (0)$ are the
ZMTL total cross sections, to be discussed below.
The function $\Phi^{\lambda} (E_R)$ accounts for the structure of the nucleus 
and is normalized to one in the ZMTL, $\Phi^{\lambda} (0)=1$. 

For $\lambda =\text{SI}$, $\Phi^{\text{SI}} (E_R ) =F^2(E_R)$ where $F(E_R)$ is the nuclear form factor.
In Eq.~(\ref{dsiggen}), therefore, the nuclear physics is separated from the particle physics.

For $\lambda =\text{SD}$,
in the standard formalism introduced by Engel in Ref.~\cite{engel}, (see \cite{engelrev,BS1,JKK} for reviews),
we have
\begin{equation}
\Phi^{SD}_E (E_R)=\frac{S(E_R)}{S(0)},
\label{Se1}
\end{equation}
with
\begin{equation}
S(E_R) = a^2_0 S_{00}(E_R) +  a_0 a_1 S_{01}(E_R) + a^2_1 S_{11}(E_R).
\label{Se2}
\end{equation}
$i,j=0,1$ are isospin indexes and $a_0$ and $a_1$ the isoscalar and isovector WIMP-nucleon scattering amplitudes written in the
isospin basis.
The  ZMTL of the functions $S_{ij}(E_R)$ is
$S_{ij}(0)\neq 1$, they are not normalized to one
and the function $S_{01}$ for some nuclei can be negative.
Particle physics and nuclear physics are not separated.

These unpleasant features of the standard formalism are avoided 
with the formalism of Divari, Kosmas, Vergados and Skouras~\cite{divari}. In this framework
we can write 
\begin{equation}
\Phi^{SD}_{V} (E_R)=\frac{\mathcal{F}(E_R)}{\mathcal{F}(0)},
\label{Sv1}
\end{equation}
with
\begin{equation}
\mathcal{F}(E_R) = a^2_0 F_{00}(E_R) + 2 a_0 a_1 F_{01}(E_R) + a^2_1 F_{11}(E_R).
\label{Sv2}
\end{equation}
Note that in this case $F_{ij}(0) =1$ by construction.
In Ref.~\cite{Cannoni} (see also \cite{divari,toivanen,vergados2010}) we have remarked
that the functions $F_{ij}(E_R)$ are practically identical
in the recoil energy interval of interest for experiments, not only 
for light nuclei but also for medium-heavy and heavy nuclei,
\begin{equation}
F_{00}(E_R)\simeq F_{01}(E_R) \simeq F_{11}(E_R).
\label{Fequiv}
\end{equation}  
Thanks to Eq.~(\ref{Fequiv}), Eq.~(\ref{Sv2}) reduces to 
\begin{eqnarray}
\Phi^{SD}_{V} (E_R) = F_{11}(E_R).
\label{Sv3}
\end{eqnarray}
Hence  the  SD ``form factor'' is determined by only one SSF. 
It does not depend anymore on the neutralino properties
as it happens in the SI scattering.

The two formalisms are equivalent and connected by 
\begin{equation}
F_{ij}(E_R)=\frac{S_{ij}(E_R)}{S_{ij}(0)}.
\label{FSequiv1}
\end{equation}
If the $S_{ij}$ are known also the $F_{ij}$ are known and vice versa.
Eq.~(\ref{Sv3}), anyway, allows for a drastic simplification of the formulas while retaining
the exact informations of nuclear shell-model calculations.
In literature, in spite of this, the formalism is largely overlooked.
In some cases  phenomenological parametrizations are used.

One example is the parametrization given in~\cite{LS,lewinsmith}
\begin{equation}
F_{\text{LS}} (q r_n)=\left\{ \begin{array}{ll}
\left(\frac{\sin(q r_{n})}{q r_{n}}\right)^2 \,&q r_n <2.55,\,q r_n>4.5,  \\	
0.047 \, &2.55 \leq \,q r_n \leq 4.5,
\end{array}\right.
\label{FSDLS}
\end{equation}
with the nuclear radius $r_n \simeq 1.0 A^{1/3}$ fm.

Another example is furnished by the parametrization implemented 
in the code $\textsf{micrOMEGAs}$~\cite{micromegas}
for the case of nuclei for which the $S_{ij}$ are not available:

\begin{equation}
F_{\text{mO}}=\frac{S_{ij}(q)}{S_{ij}(0)}=\exp\left(-\frac{q^2 R_A^2}{4}\right),  
\label{FSDmicr}
\end{equation}
where $R_A = 1.7 A^{1/3} -0.28 -0.78
(A^{1/3}- 3.8 
+ [(A^{1/3} - 3.8)^2 + 0.2]^{1/2})$.
These expressions are used also in recent literature~\cite{simple,shan,an} even in the case that
the functions $S_{ij}$ or $F_{ij}$ are known.
It is thus interesting to compare them with $F_{11}$.

Figure~\ref{fig1:SSF} shows the normalized SSF $F_{11}$, 
$F_{\text{LS}}$ and $F_{\text{mO}}$ for one light nucleus,
\fluorine, one medium-heavy, \germanium, and one heavy nucleus, \iodine,
all of them largely employed in current experiments. 
The function $F_{11}$ for \fluorine ~is taken from Ref.~\cite{divari},
for $^{73}$Ge is obtained from the function $S_{11}$ of Ref.~\cite{dimitrov},
for \iodine ~from the function $S_{11}$ of Ref.~\cite{ressell} (set calculated with the Bonn A potential). 

In the abscissas we use the dimensionless variable $y=(qb/2)^2$ where $b=1\,\text{fm}\,A^{1/6}$
is the oscillator size parameter. This variable is the natural one 
employed in shell-model calculations using harmonic oscillator wave functions.
The functional form of $S_{ij}$ and $F_{ij}$ is typically a polynomial or a polynomial times an exponential
in $y$ or $u=2y$~\cite{BS1,divari}. 
The recoil energy is easily found to be related to $y$ by $E_R =80\times y\times A^{-4/3}$ MeV.
For clearness we report also the corresponding 
recoil energies for each nucleus on a second abscissa. 
The interval $0<y<1$ covers the recoil energies interval accessible experimentally but
in the case of fluorine the relevant region is only up to $y\sim 0.1$.   

The approximation furnished by $F_{\text{LS}}$ is reasonable both at low recoil energies and
at higher energies in the region of the plateau, especially for the heavy nucleus.
This is not surprising, for this parametrization was introduced~\cite{LS} to fit the 
SSF in Xe and Nb~\cite{engel,engel2}.
The approximation furnished by $F_{\text{mO}}$ is much worse in all the cases.
A different Gaussian parametrization is given for example in Refs.~\cite{gondoloDD,bertonereview}.

We stress again that for the nuclei for which 
the functions $F_{ij}$ or $S_{ij}$ have been published, 
there is no need of phenomenological fits or parametrization.
The normalized function $F_{11}$ accounts for the results of the most accurate 
spin structure function calculations and at the same time allows to separate 
the nuclear physics from the particle physics in SD the cross section.

\subsection{Differential and total event rate}
\label{sec2.2}

In SD scattering, given the  
neutralino-proton and neutralino-neutron cross sections 
$\sigma^{SD}_{p,n} =(\mu^2_p /\pi) 3|a_{p,n}|^2$, the total cross section at $q=0$ reads
\begin{equation}
\sigma_{{A}}^{\text{SD}}(0) = \left( \frac{\mu_A}{\mu_p} \right)^2 \frac{1}{3}
\left({\Omega}_p(0)  \sqrt{ \sigma^{\text{SD}}_p}
+ \varrho {{\Omega}_n (0)} \sqrt{{\sigma^{\text{SD}}_n}}\right)^2.
\label{sigSD}
\end{equation}
$\mu_p$ is WIMP-proton reduced mass and
\begin{equation}
{\Omega}_{p,n}(0)  = 2\sqrt{\frac{J+1}{J}} \langle \mathbf{S}_{p,n} \rangle,
\label{opn}
\end{equation}
are the spin matrix elements of the proton and neutron groups.
We remind that $\langle \mathbf{S}_{p,n} \rangle \equiv  \langle J,M_J =J| S^z_{p,n}|J,M_J =J  \rangle$.
In general both  the SD WIMP-nucleon scattering amplitudes $a_p$ and $a_n$ 
($a_{p,n}=(a_0 \pm a_1)/2$) and the nuclear matrix elements can have opposite sign, 
hence $\varrho=\pm1$ is the relative sign between 
$|{\Omega}_p (0)  {a_p }|$ and $|{\Omega}_n(0)  {a_n }|$.
An \textit{ab initio} derivation of the SD cross sections using the formalism 
of Ref.~\cite{divari} is given in the Appendix.

In the SI case, for the neutralino we have $\sigma^{\text{SI}}_p \simeq \sigma^{\text{SI}}_n \equiv \sigma^{\text{SI}}$,
the standard total cross section at $q=0$ is
\begin{equation}
\sigma_{{A}}^{\text{SI}}(0) = \left(\frac{\mu_A}{\mu_p} \right)^2 A^2 \sigma^{\text{SI}}.
\end{equation} 

The differential recoil rate is obtained by
folding Eq.~(\ref{dsiggen}) with the velocity distribution function. We use the 
standard truncated Maxwellian~\cite{lewinsmith}: 
\begin{eqnarray}
f_1 (v)&=&\frac{v}{v_0 v_E} f(v),\\
f(v)& =&\frac{1} {\kappa }\left[\exp\left({-\frac{(v-v_E)^2}{v^2_0}}\right)-\exp{\left(-\frac{(v+v_E)^2}{v^2_0}\right)}\right],\nonumber\\
\kappa&=&\sqrt{\pi} \text{erf}(z)-2z \exp({-z^2}),\;\; z=\frac{v_{esc}}{v_0}.\nonumber
\end{eqnarray}
$v_{esc}$ is the escape velocity, $v_0$ the velocity of the Sun
$v_E$ the velocity of the Earth.

Taking  $\rho_0 =0.3$ GeV/cm$^{3}$ as the local dark matter density density,
$\epsilon_0 = 2 \mu_A v^2_0 ({\mu_A}/{m_A})$ the typical recoil energy and 
$\Phi^{\text{SI}} = F^2 (E_R)$,  $\Phi^{\text{SD}} = F_{11} (E_R)$, we can write
\begin{eqnarray}
\frac{dR^{\lambda}}{dE_R}&=&\frac{\rho_0 v_0 }{m_\chi m_A }\sigma_{A}^{\lambda}(0)\frac{dt^{\lambda}}{dE_R},\\
\frac{dt^{\lambda}}{dE_R}&=& \frac{\Phi^{\lambda}(E_R)}{\epsilon_0}
\int_{v_{\min}(E_R)}^{v_{\max}} \frac{dv}{v_E} f(v).
\label{tSD}
\end{eqnarray}
The total rate is simply given by 
\begin{eqnarray}
R^{\lambda}&=&\frac{\rho_0 v_0 }{m_\chi m_A } 
\sigma_{A}^{\lambda}(0) t^{\lambda},\label{rateSD}\\
t^{\lambda} &=&\int_{E_1}^{E_2} {dE_R}
\frac{dt^{\lambda}}{dE_R}.
\end{eqnarray}
The integration limits are $v_{\min}(E_R) = v_0 \sqrt{{E_R}/{\epsilon_0}}$, $v_{\max} = v_{esc}$,
$E_1 =E_{th}$, $E_2 = \min(E^{exp}_2 , E_{\max})$ where the maximal recoil energy is 
$E_{\max}=\epsilon_0 \left({v_{\max}}/{v_0}\right)^2$.
The energy threshold $E_{th}$ and $E^{exp}_2$ give the energy interval chosen by an experiment to analyze  the data.
For comparison with a given experiment using
specified nuclei and detection methods, 
if necessary, one should account in the previous formulas for the energy resolution and efficiencies that  may depend on the energy. 

In the following we use the Helm form factor in the 
parametrization proposed in Ref.~\cite{lewinsmith}: 
\begin{eqnarray}
&&F^2 (q)=\left(3\frac{j_1 (q r_n )}{q r_n}\right)^2 \exp{(-q^2 s^2)},\label{FF}\\
&&j_1 (x)=\frac{\sin x}{x^2} - \frac{\cos x}{x},\;\;
r_n = \sqrt{c^2 +\frac{7}{3}\pi^2 a^2 -5s^2}\; \,\text{fm}, \nonumber\\
&&s=0.9 \,\text{fm},\;\;
a=0.52 \,\text{fm},\;\;
c=(1.23 A^{1/3}- 0.6)\, \text{fm}.\nonumber
\end{eqnarray}
In literature one can find other parametrization~\cite{duda},
or form factors obtained directly by shell-model calculations~\cite{KV,divari,toivanen}. 
We use here Eq.~(\ref{FF}) because employed practically by all the experimental groups.

\section{Model independent upper limits}
\label{sec3}

As an application of the previous formalism we discuss the so-called
model independent method for setting upper limits on neutralino 
cross sections and give an alternative proof of Eq.~(13) of Ref.~\cite{tovey}.  

Let us consider a nucleus such that the SI rate is negligible compared 
to SD one: in supersymmetric models the SD 
rate roughly dominate in nuclei with mass number $A\leq 20$ while SI dominate at larger 
mass numbers due to $A^2$ proportionality~\cite{JKK}. We return on this
point in the next section.

We introduce the factors 
\begin{eqnarray}
\phi_A &=&\frac{\rho_0 v_0}{m_{\chi} m_{A}}, 
\label{fi}\\
\mathcal{C}^{p,n}_A &= & \frac{\mu_A}{\mu_p}\frac{{\Omega}_{p,n}(0)}{\sqrt{3}}.
\label{cfac}
\end{eqnarray}
Eq.~(\ref{rateSD}), with the aid of Eqs.~(\ref{sigSD}),~(\ref{fi}),~(\ref{cfac}), 
thus becomes 
\begin{equation}
R^{\text{SD}}= \phi_A \left(\mathcal{C}^{p}_A \sqrt{ \sigma^{\text{SD}}_p} \pm  
\mathcal{C}^{n}_A \sqrt{ \sigma^{\text{SD}}_n}\right)^2  t^{\text{\text{SD}}}_A .
\end{equation}
If an experiment with exposure $\mathcal{E}_{A} = M_A \times T$, 
($M_A$ is the mass fraction of the element with mass number $A$
and $T$ the time of live data taking)
have no statistically significant evidence, 
then an upper limit at some confidence level is put on the number of events $N^{UL}$.   
For each unknown $m_\chi$ this is converted in an upper limit on the 
cross section requiring $R \times \mathcal{E} < N^{UL}$, that is

\begin{equation}
\left(\mathcal{C}^{p}_A \sqrt{ \sigma^{\text{SD}}_p} \pm  
\mathcal{C}^{n}_A \sqrt{ \sigma^{\text{SD}}_n}\right)^2 < \frac{N^{UL}}
{\phi_A  t^{\text{\text{SD}}}_A \mathcal{E}_A}.
\label{sigUL}
\end{equation}
The right-hand side of (\ref{sigUL}) is by definition the experimental 
upper limit on the neutralino-nucleus SD cross section,
let us call $\sigma^{\text{lim}}_A$ as in Ref.~\cite{tovey},
\begin{equation}
\sigma^{\text{lim}}_A \equiv \frac{N^{UL}}
{ \phi_A  t^{\text{SD}}_A \mathcal{E}_A}.
\label{sigA}
\end{equation}
Furthermore, utilizing the same name of Ref.~\cite{tovey}, we define the quantities
\begin{equation}
\sigma^{\text{lim(A)}}_{p,n}\equiv\frac{\sigma^{\text{lim}}_A}{(\mathcal{C}^{p,n}_A)^2}.
\label{pnsiglim}
\end{equation}
Dividing both members of (\ref{sigUL}) by (\ref{sigA}) and using the quantities
(\ref{pnsiglim}) we  arrive at
\begin{eqnarray}
\left(\frac{\sqrt{\sigma^{\text{SD}}_p}}{\sqrt{\sigma^{\text{lim(A)}}_{p}}}\pm
\frac{\sqrt{\sigma^{\text{SD}}_{n}}}{\sqrt{\sigma^{\text{lim(A)}}_{n}}}\right)^2 <1 ,
\label{diseg}
\end{eqnarray}
that is exactly Eq.~(13) of Ref.~\cite{tovey} in the case of the allowed region in 
($\sigma_p , \sigma_n$) plane.

\begin{figure*}[t!]
\includegraphics*[scale=0.42]{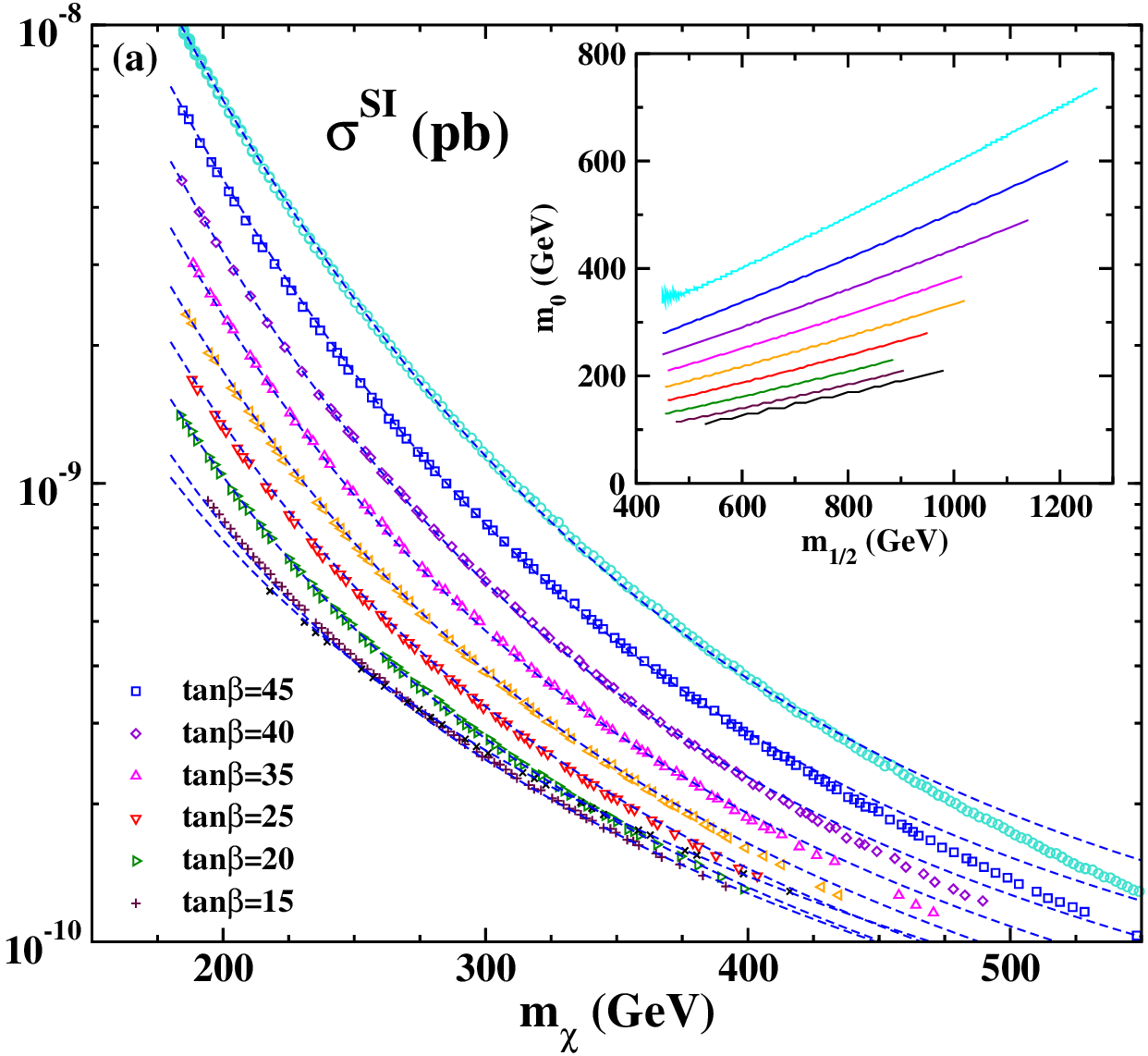}
\includegraphics*[scale=0.42]{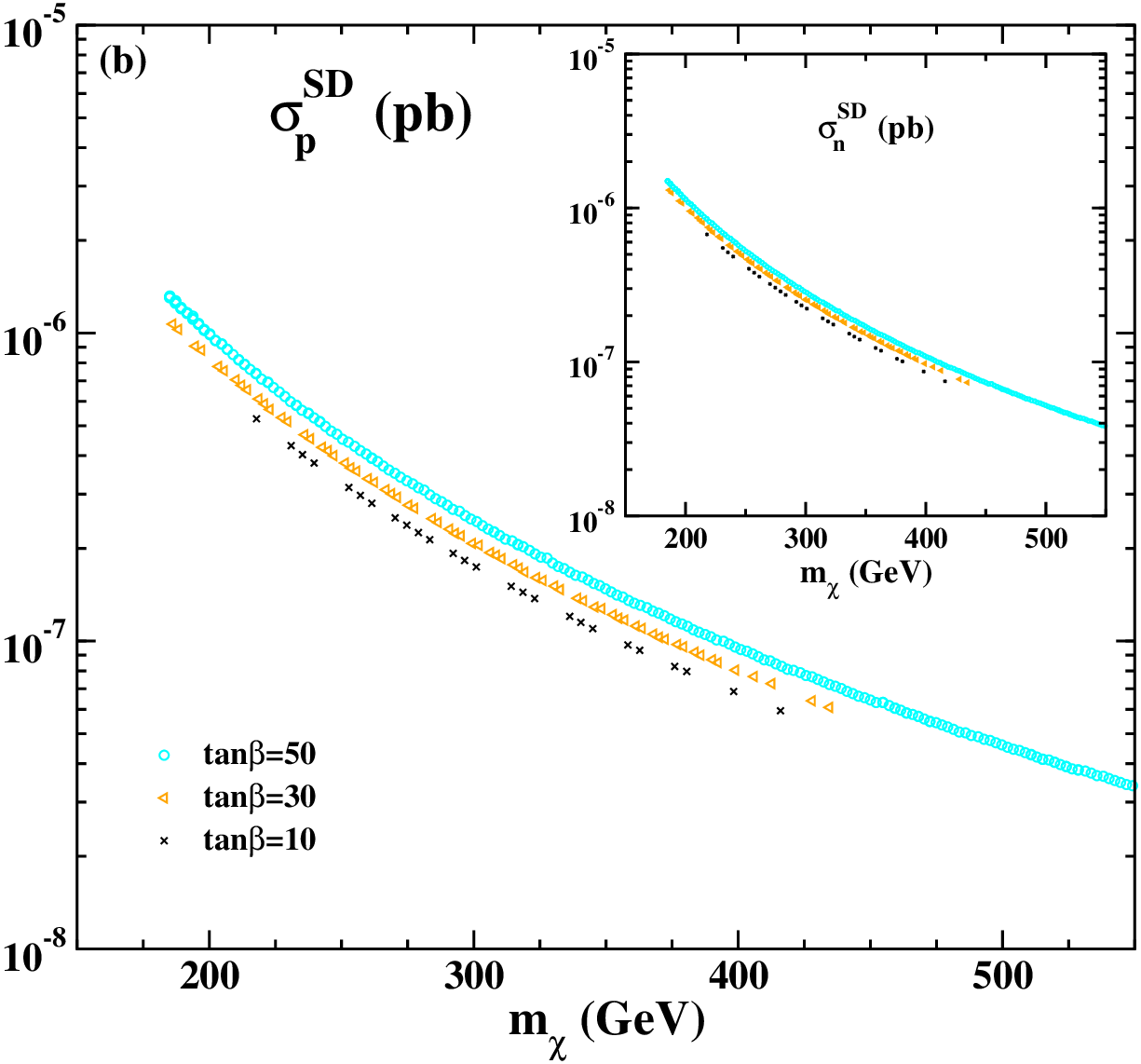}
\caption{(a) Spin-independent neutralino-nucleon cross section 
in the stau co-annihilation region with $A_0 =0$ and $\tan\beta$ from 10 
to 50 in step of five as a function of the neutralino mass. 
The dashed line is obtained with the fitting formula, Eq.~(\ref{fit}),
with coefficients given in Table~\ref{table}.
The insert shows for each value of $\tan\beta$ the strip in
the plane ($m_{1/2}$, $m_0$) allowed by WMAP constraints on the 
relic density $0.096<\Omega h^2 < 0.128$ and satisfy accelerator constraints.
(b) Spin-dependent neutralino-proton cross section and, in the insert, the spin-dependent neutralino-neutron 
cross section for $\tan\beta =10$, 30 and 50.  }
\label{fig2:SI--SD}
\end{figure*}

In Ref.~\cite{tovey} the nucleon cross section limits in Eq.~(\ref{pnsiglim}) are defined 
as basic quantities that  then are combined to give Eq.~(\ref{diseg}).
To do this it is necessary to assume that for a given nucleus it is possible to set 
separately limits on the SD-proton and SD-neutron cross sections even in the case that 
one contribution is clearly sub-dominant.
These assumptions and the method were criticized in in Refs.~\cite{bednyakov_aspects,BK,BS1}. 

In reality our derivation shows that such hypothesis are unnecessary and that the full justification 
of Eq.~(\ref{diseg}) only relies on the factorization of the particle physics from nuclear physics degrees of freedom
and has a general validity\footnote{This result was also implicitly obtained, with 
different notations and considering the case of general phases, in Ref.~\cite{vergados2010}.}.

Another common misunderstanding about Eq.~(\ref{diseg})
is that it is based on ZMTL total cross section and that it
does not take into account
the exact momentum dependent structure function.

Actually, we see that using $F_{11}$, the correct behavior of the  SSF can
be taken into account in the upper limit $\sigma^{\text{lim(A)}}$ by the factor $t^{\text{SD}}$,
see Eq.~(\ref{sigA}).

On the other hand, the ``upper limits'' on the single proton or
neutron cross sections, Eq.~(\ref{pnsiglim}), are just useful quantities
introduced to write Eq.~(\ref{sigUL}) in the compact form (\ref{diseg}). 
They become the actual experimental upper limits if, 
for the nucleus from which these are determined 
and in a specific WIMP model, one can prove that the protons 
contribution is dominant over the neutrons contribution or vice-versa (given the dominance
of the SD rate over the SI rate). 
In general the exclusion curves on the single cross sections
are fundamentally  indicative of the experiments sensitivity and 
cannot constrain particle physics models in a universal way.

\section{SD scattering and  the $\bm{\tilde{\tau}}_{\text{CR}}$ }
\label{sec4}

To further clarify the last  point, we choose a specific particle physics model, that is the 
constrained minimal supersymmetric standard model (CMSSM) with R-parity conservation.
We consider the parameter space with fixed trilinear scalar coupling $A_0 =0$,
positive Higgs mixing term ($\mu>0$) which is the benchmark supersymmetric 
theory for phenomenological and experimental studies~\cite{ellisWMAP}. If the neutralino is
required to furnish the cosmological relic density inferred by WMAP~\cite{wmap}, then, 
for fixed $\tan\beta$ only specific regions in the $(m_{1/2}, m_0)$ plane are
left. In the ($\tilde{\tau}_{\text CR}$)
the lightest stau is almost degenerate in mass with the neutralino and the 
co-annihilation of the two particles in the early Universe brings the  value
of the relic density in the favored WMAP interval. 
This parameter space is still untouched by direct detection
experiments and LHC just started to explore it~\cite{ellisfit,farina,bertone}; moreover it 
will be hard to probe it
with indirect detection methods such as $\gamma$-ray from neutralino annihilation in the 
halos~\cite{CannoniIB,masc,ellisgammas}.

The strips in the the plane ($m_{1/2}$, $m_0$)~\cite{ellisWMAP} for varying 
$\tan\beta$ from 10 to 50 in step of 5,  are shown in the 
insert of Fig.~\ref{fig2:SI--SD}(a).
The strips and the cross sections are obtained with $\textsf{DarkSUSY}$~\cite{darksusy}, 
imposing WMAP constraints on the relic density $0.096<\Omega h^2 < 0.128$, 
accelerator constraints on the lightest Higgs, $m_h >114$ GeV and 
chargino mass $m_{\chi^{+}_1}>103.5$ GeV and the flavor physics constraint from 
bottom quark radiative transitions.

In the same figure the SI neutralino-proton cross section as a function of the neutralino mass is shown. The 
SD neutralino-proton and neutralino-neutron cross section are shown in 
Fig.~\ref{fig2:SI--SD}(b) and in the insert of Fig.~\ref{fig2:SI--SD}(b), respectively.
Two general features are worth noting: the SI cross section depends on $\tan\beta$ more strongly than
the SD cross sections, the former varying by an order of magnitude and the latter
by a factor less than 2; the SD are  $\mathcal{O}(10^2)$ larger than the SI, in agreement with~\cite{ellisUncertain}. 

The neutralino field in the mass basis can be written as 
$\chi^0_1 \equiv N_{11}\tilde{B}+N_{12}\tilde{W}^0 +N_{13}\tilde{H}_1 + N_{14} \tilde{H}_2$,
where $N_{1i}$ are the elements of the matrix that diagonalizes the neutralino mass matrix,
$\tilde{B}$, $\tilde{W}^0$ are the neutral gaugino fields and $\tilde{H}_1$, $\tilde{H}_2$
the neutral higgsino fields. 
In all the  considered parameter space the neutralino is  
bino--like: we find numerically
$N_{11}\sim 0.99 \gg N_{13} \sim 10^{-3} \gg N_{12},N_{14}$. This means that the coupling 
to the $Z$ boson that is driven by the higgsinos couplings proportional to $N_{13}$ and $N_{14}$
is heavily suppressed; the cross section is determined by squarks exchange.
Analogously also in the SI case to the CP--even Higgs $h$ and $H$  are suppressed
by $N_{13}$ and $N_{14}$ and the cross sections is thus mainly determined by squark exchange.
Anyway, the couplings of the Higgs to down-type quark become $(\tan\beta)^2$ enhanced at large
$\tan\beta$. The two contributions thus can be of the same order and SI cross section
is more sensitive to variations of $\tan\beta$.

\subsection{$^{\bm{19}}$F, $^{\bm{127}}$I   and the $\bm{\tilde{\tau}}_{\text{CR}}$ }
\label{sec4.1}

To discuss the relation between the SI and SD rates we consider the light nucleus  \fluorine~that is 
known to furnish the best sensitivity to the proton SD cross section~\cite{ellisflores,pacheco,divari}
and  \iodine, that have both good SI and SD sensitivities.

For \fluorine ~we use the spin matrix elements of Ref.~\cite{divari} that give
${\Omega}^{19}_p(0) =1.646$ and ${\Omega}^{19}_n(0) = -0.030$.
In this case the neutrons contribution in the SD rate 
can be safely neglected. We remark  that the first nuclear shell-model calculation for \fluorine~\cite{pacheco}
found $\langle \mathbf{S}_{p} \rangle = 0.441$, $\langle \mathbf{S}_{n} \rangle =-0.109 $.
The successive  calculation of Ref.~\cite{divari} using a more realistic interaction,  found 
$\langle \mathbf{S}_{p} \rangle = 0.4751$ and $\langle \mathbf{S}_{n} \rangle = -0.0087$.
The  protons contribution is thus similar but the neutrons contribution
is clearly negligible. The statement 
that the neutrons contribution is relevant, see for example~\cite{simple}, in light of the more accurate 
calculation of Ref.~\cite{divari}, is doubtful.

\begin{figure*}[t!]
\includegraphics*[scale=0.67]{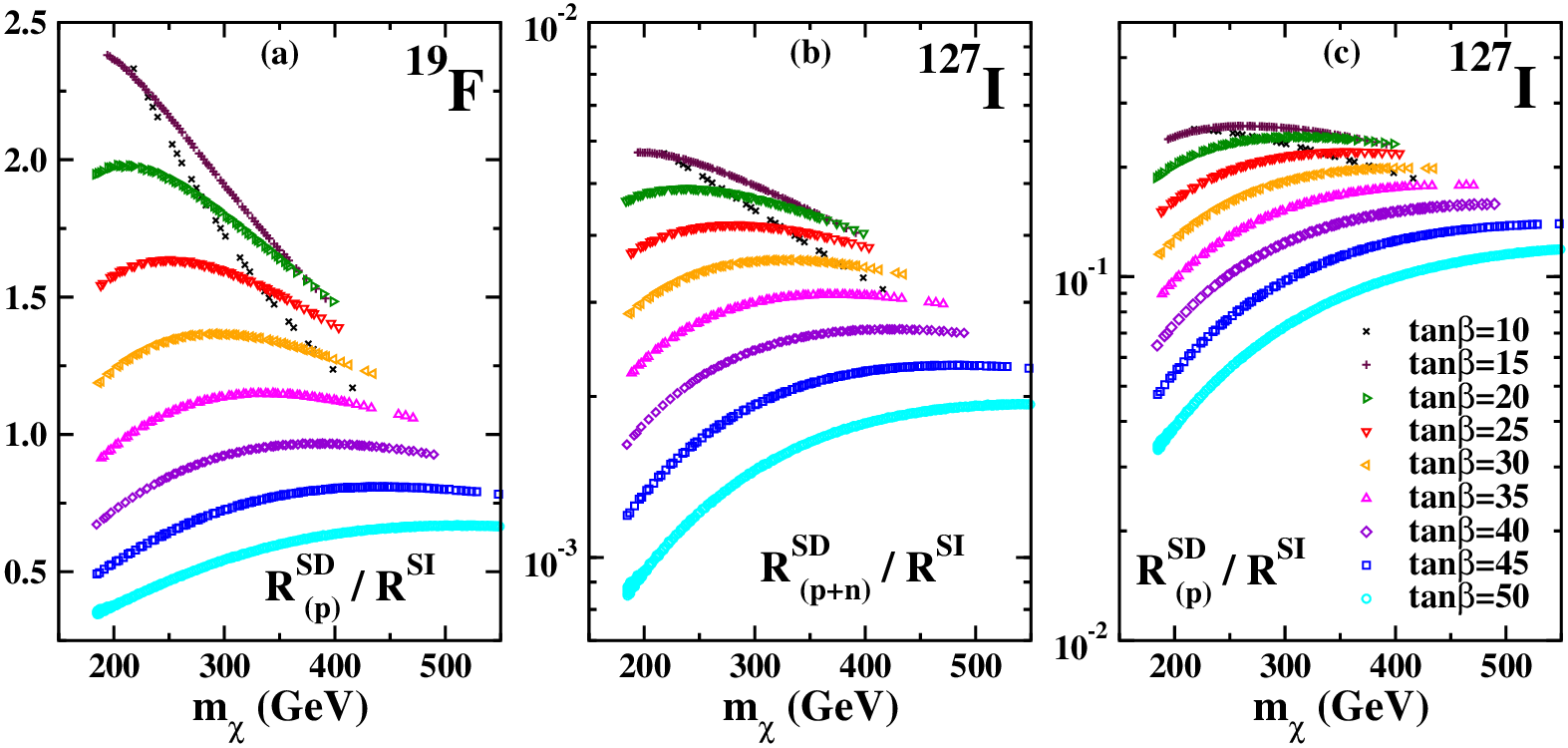}
\caption{Ratio of the spin-dependent total event rate over the spin-independent rate $R^{\text{SD}} /R^{SI}$
varying $\tan\beta$ in the stau co-annihilation region of the CMSSM with $A_0 =0$ and $\mu>0$.
In panel (a) for \fluorine.
In panel (b), the ratio is plotted for \iodine taking into account both the protons 
and neutrons contribution in the spin-dependent rate; in panel (c) only the proton contribution is included.
The points of the parameter space are the same as in Fig.~\ref{fig2:SI--SD}.}
\label{fig3:ratios}
\end{figure*}

As reminded above, for light nuclei like fluorine the SD rate can be dominant over the SI,
but this has to be checked in each particular WIMP model.
We show the ratio $R^{\text{SD}} /R^{SI}$ for fluorine in Fig.~\ref{fig3:ratios}(a). 
The SD rate is bigger by a factor 
up more than 2 at low and medium $\tan\beta$ but it is smaller than the SI 
rate at large $\tan\beta$; in any case the two rates are always of the same 
order of magnitude. 
The SI rate cannot be completely neglected at high $\tan\beta$ and  for 
lower $\tan\beta$, neglecting it, one underestimates the total rate 
(see Ref.~\cite{bedrates} for the case of general MSSM). %
The exclusion plots in the 
($m_\chi$, $\sigma^{\text{SD}}_p$) are inaccurate for the $\tilde{\tau}_{\text{CR}}$. 
In this case one has to draw an exclusion plot in the 
($\sigma^{\text{SD}}_p$, $\sigma^{SI}$) plane for each fixed mass,
the so-called mixed coupling approach~\cite{bernabei}.

Nuclear shell model calculations give for \iodine ~${\Omega}^{127}_p(0) =0.731$ and ${\Omega}^{127}_n(0) =0.177$
(spin matrix elements obtained with the 
potential Bonn-A from Ref.~\cite{ressell}).
Although proton favoring, the neutrons group contribution to the nuclear spin
is of the same order of magnitude. 
If the neutralino couplings to the proton and neutron are similar, the neutrons contribution to the 
nuclear spin must be considered. This indeed is what happens in the \scr ~where
$0.75< \sigma^{\text{SD}}_p/\sigma^{\text{SD}}_n <0.9$~\cite{Cannoni} for $\tan\beta$ 
between 10 and 50. Furthermore, $a_p <0$ and $a_n >0$ thus a cancellation in the SD rate 
is expected because the products $a_p \langle\mathbf{S}_{p}\rangle$ and $a_n \langle\mathbf{S}_{n}\rangle$ are of the 
same order and have opposite sign.
Fig.~\ref{fig3:ratios}(c) shows the ratio of $R^{\text{SD}} /R^{SI}$ in \iodine
~only considering the proton contribution, while in Fig.~\ref{fig3:ratios}(b) both  are included.
Due to the $A^2$ proportionality, the SI rate always dominate by a factor from 4 to 25 in Fig~\ref{fig3:ratios}(c),  
but the  cancellation  makes the SD rate from 2 to 3 orders of magnitude 
smaller than the SI, Fig~\ref{fig3:ratios}(b).

In the case of $\tilde{\tau}_{\text{CR}}$, hence, iodine can only constrain the SI interaction.
The exclusion plots in the planes ($m_\chi$, $\sigma^{\text{SD}}_p$),
($m_\chi$, $\sigma^{\text{SD}}_n$) or the combined 
($\sigma^{\text{SD}}_p$, $\sigma^{\text{SD}}_n$) at fixed neutralino mass, derived 
using \iodine ~cannot constrain the $\tilde{\tau}_{\text{CR}}$, for they are derived 
neglecting the dominant SI contribution or the equally important neutrons contribution
that almost cancel the protons one.

\subsection{COUPP and the  $\bm{\tilde{\tau}}_{\text{CR}}$}
\label{sec4.2}

\begin{figure*}[t!]
\includegraphics*[scale=0.63]{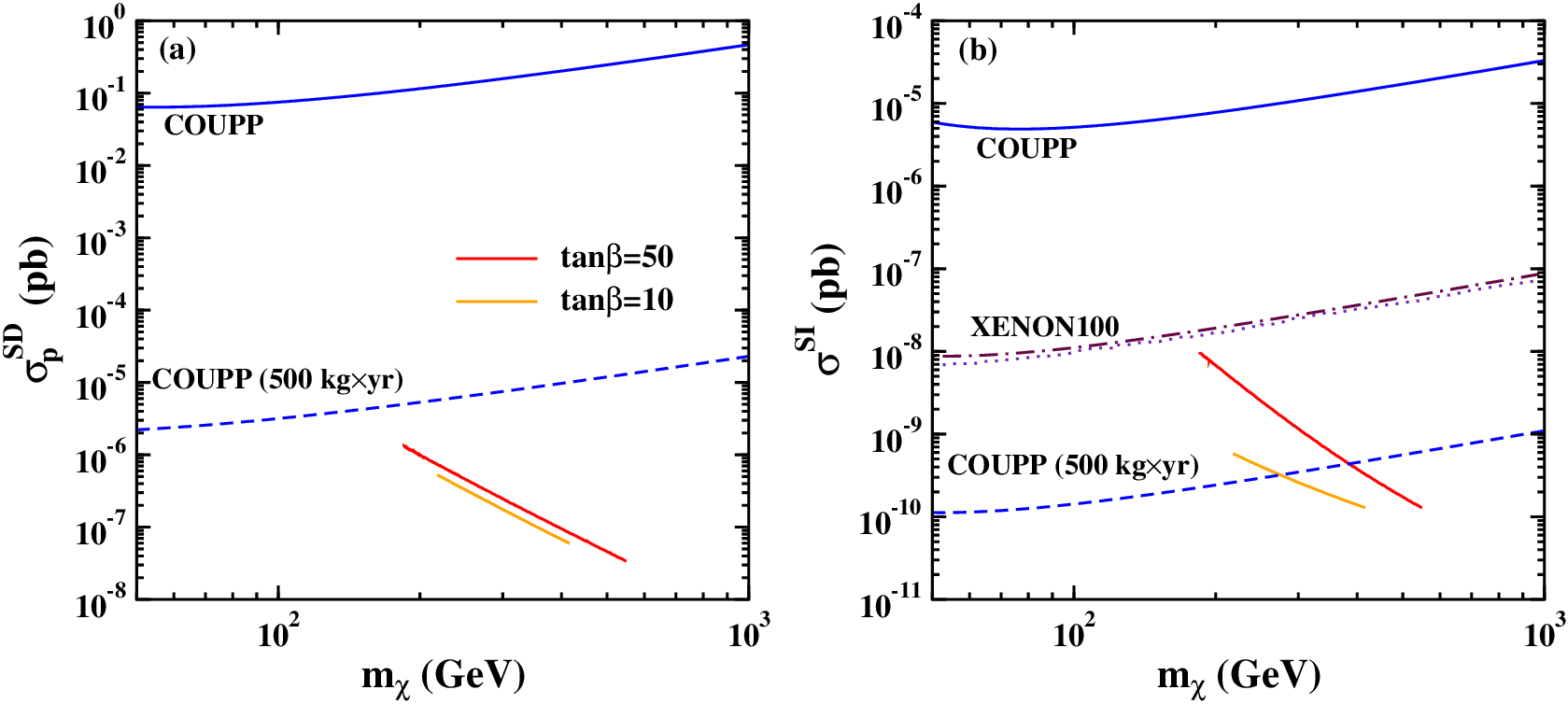}
\caption{
(a) Spin-dependent neutralino-proton cross sections for $\tan\beta =10$ (orange line)
and 50 (red line) in the stau co-annihilation region with $A_0 =0$, $\mu>0$. The blue
solid line represent the present upper limit from COUPP, the blue-dashed line
the extrapolated COUPP limit with an effective exposure of 500 kg$\times$ year and threshold 
at 7 keV.
(b) The same as (a) but for the proton spin-independent cross section. 
The dashed-dotted line is the upper limit from  XENON100 calculated as explained in the text,
the dotted line the limit published in Ref.~\cite{xenon100}.}
\label{fig4:COUPP}
\end{figure*}

The two nuclei discussed so far are the detecting medium of 
COUPP~\cite{coupp}, a bubble chamber with CF$_3$I.
We use the latest data from Ref.~\cite{coupp}: an affective exposure of CF$_3$I 
after cuts of $\mathcal{E}=28.1$ kg$\times$days, 50$\%$ efficiency, $E_{th}=21$ keV, $N^{UL}=6.7$ at 
90$\%$ confidence level and the same values of the velocities,
$v_0 =230$ km/s, $v_{esc} =650$ km/s and an average velocity of the Earth $v_E = 244$ km/s.
In Fig.~\ref{fig4:COUPP}(a) the blue solid line is the present limit on SD WIMP-proton cross 
section  derived from the fluorine fraction, while in  Fig.~\ref{fig4:COUPP}(b)
is the limit on the SI cross section derived 
from the iodine fraction. 
The blue dashed lines are limits extrapolated with the same $N^{UL}$, 100$\% $ efficiency,
effective exposure 500 kg$\times$yr and threshold at 7 keV. 
The red solid lines are the cross sections for $\tan\beta=50$, the orange ones for $\tan\beta=10$.

The indication that we derive from Fig.~\ref{fig4:COUPP} is that it will be unlikely for COUPP
to probe the \scr ~by SD scattering
unless very large exposures of fluorine
are achieved. On the other hand, a part of the parameter space will be probed by the SI scattering with iodine. 
This is not a limitation for the \scr  ~since the two cross sections are clearly correlated.
A constraint on $\sigma^{\text{SI}}$ automatically implies a constraint on 
$\sigma^{\text{SD}}$. As a matter of fact, should evidence be reported by more experiments and rates 
measured in fluorine, iodine and other elements like xenon, argon or  germanium, the
the full information on the SD sector can be reconstructed~\cite{Cannoni,pato}.

As discussed in Section~\ref{sec4.1}, the SI rate in the \scr ~cannot be neglected for fluorine.
Considering a mixed SI--SD approach with a fixed neutralino mass we obtain:
\begin{equation}
\sigma^{\text{SD}}_p < \sigma^{\lim(F)}_p-\frac{(\mathcal{C}^{SI}_F)^2 t^{SI}_F}
{(\mathcal{C}^{p}_F)^2 t^{\text{SD}}_F} \sigma^{SI}.
\end{equation}
In analogy with Section~\ref{sec2} we set $\mathcal{C}^{SI} =(\mu_A /\mu_p) A$.
At $\tan\beta =50$, where the SI rate is more important,
the correction term on the right-hand side gets values larger than $\sigma^{\text{SD}}_p$,
in any case the largest values are of order $10^{-6}$ pb. 
These values when compared with the present limits, $\sigma^{\lim(F)}_p \simeq 10^{-1}$ pb from COUPP 
and $\simeq 10^{-2}$ pb form SIMPLE~\cite{simple}, are anyway negligible.
Hence one should start to consider the SI rate only when the exposure
is such that the sensitivity reaches the values of $\sigma^{\text{SD}}_p$ predicted by the
model.

\section{Constraining the ($\bm{\text{m}}_{\bm{1/2}}$, $\bm{\text{tan}}$$\bm{\beta}$) plane }
\label{sec5}

\begin{table*}[htbp!]
\caption{Coefficients for the fitting formula of Eq.~(\ref{fit}).} 
\label{table}
\begin{ruledtabular}
\begin{tabular}{cccccc}
 $k$ & $(\sigma)_{k0}$ (pb) & $(\sigma)_{k1}$ (pb) & $(\sigma)_{k2}$ (pb) & $(\sigma)_{k3}$ (pb)
& $(\sigma)_{k4}$ (pb)\\
\hline
$2 $   & 2.469$\times 10^{-9}$ & 5.085$\times 10^{-11}$ & 5.432$\times 10^{-12}$ 
& 1.783$\times 10^{-13}$ & -6.089$\times 10^{-16}$ \\
$3$   & 2.716$\times 10^{-9}$ & -9.790$\times 10^{-10}$ & 3.92$\times 10^{-11}$ 
& -6.413$\times 10^{-13}$ & -6.059$\times 10^{-15}$ \\
$4$   & 1.395$\times 10^{-8}$ & -2.029$\times 10^{-9}$ & 2.143$\times 10^{-10}$ 
&  -6.711$\times 10^{-12}$ & 9.481$\times 10^{-13}$ \\
\end{tabular}
\end{ruledtabular}
\end{table*}

In Fig.~\ref{fig4:COUPP}(b) we also show for comparison the present upper limit 
of XENON100~\cite{xenon100}, which is the most stringent on the SI cross section.
To be consistent  with COUPP,  we have calculated the XENON100 plot using the same values
of the velocities given above and the following data: effective exposure of $1471$ kg$\times$days,
energy threshold at 8.4 keV and $N^{UL}=5.62$ at 90$\%$ confidence level deduced by the 
Feldman-Cousins method~\cite{fc} with 3 events observed and mean background 1.8.
As for the COUPP limits, we have calculated this curve using the total event rate 
without energy resolution function.
Our curve differs by few percent from published one, dotted line in Fig.~\ref{fig4:COUPP}(b).
The latter is obtained with
a statistical analysis of the energy spectrum that take into account all the experimental
uncertainties and with values of the velocities
$v_0 =220$ km/s, $v_{esc} =544$ km/s and  $v_E = 232$ km/s.

This exercise shows that for masses above 50 GeV the limits are more robust and less 
sensitive to the experimental details, statistical method to analyse the data and velocities
(needless to say this is not true in the low mass region). 
In the high mass range $m_\chi >50$ GeV the exclusion limits are also robust 
against changes of the velocity distributions~\cite{mccabe}, being the major source of
uncertainty a factor of two in $\rho_0$.

Since we have remarked above that the \scr ~will be probably probed through the SI scattering,
we further investigate what kind of information on the \scr ~parameter space can be extracted.
We note from Fig.~\ref{fig2:SI--SD}(a) that the SI cross section 
is a smooth decreasing function of the neutralino mass
when $m_0$ and $m_{1/2}$ are varied along the WMAP allowed lines for fixed $\tan\beta$.
Clearly it is also a continuous function of this parameter. Therefore we can look for a general
fitting formula valid for all the values of $\tan\beta$. 
We first fit each $\sigma^{SI}$ of Fig.~\ref{fig2:SI--SD}(a) for a given value of $\tan\beta$ 
with the function 
\begin{equation}
\sigma =\sum\limits_{k=2}^{4} \varsigma_k \left(\frac{100 \text{ GeV}}{m_{\chi}}\right)^k,
\end{equation}
and than the coefficients $\varsigma_k$ are fitted with a 4th order polynomial
in $\tan\beta$. We thus find 
\begin{equation}
\sigma^{SI}(\tan\beta, m_\chi )=\sum\limits_{k=2}^{4}
\left[
\sum\limits_{i=0}^{4} \sigma_{ki}\,(\tan\beta)^i 
\left(\frac{100\text{ GeV}}{m_\chi}\right)^k 
\right].
\label{fit}
\end{equation}
The coefficients of the fit $\sigma_{ki}$ are given in Table~\ref{table}.
The fit obtained with Eq.~(\ref{fit}) is shown in Fig.~\ref{fig2:SI--SD}(a) with a dashed line.

Analogously, the neutralino mass can be  parametrized along the WMAP lines. We find 
that for all the values of $\tan\beta$ it holds
\begin{equation}
m_\chi \simeq 0.44\; m_{1/2} -15\text{ GeV}.
\label{massfit}
\end{equation}
While the  slope 0.44 is found for all the values, the constant negative term is an average value,
since  it presents a very mild dependence on $\tan\beta$
that anyway is not important for what follows. 
Thus using Eq.~(\ref{massfit}) in Eq.~(\ref{fit}) we end up with 
a formula $\sigma^{SI}(\tan\beta, m_{1/2})$ for  the cross section in terms of the 
fundamental parameters $m_{1/2}$ and $\tan\beta$. In last analysis,
this allows to convert an upper limit on the event rate directly into an exclusion 
plot in the ($m_{1/2}$, $\tan\beta$) plane.

\begin{figure}[t!]
\includegraphics*[scale=0.4]{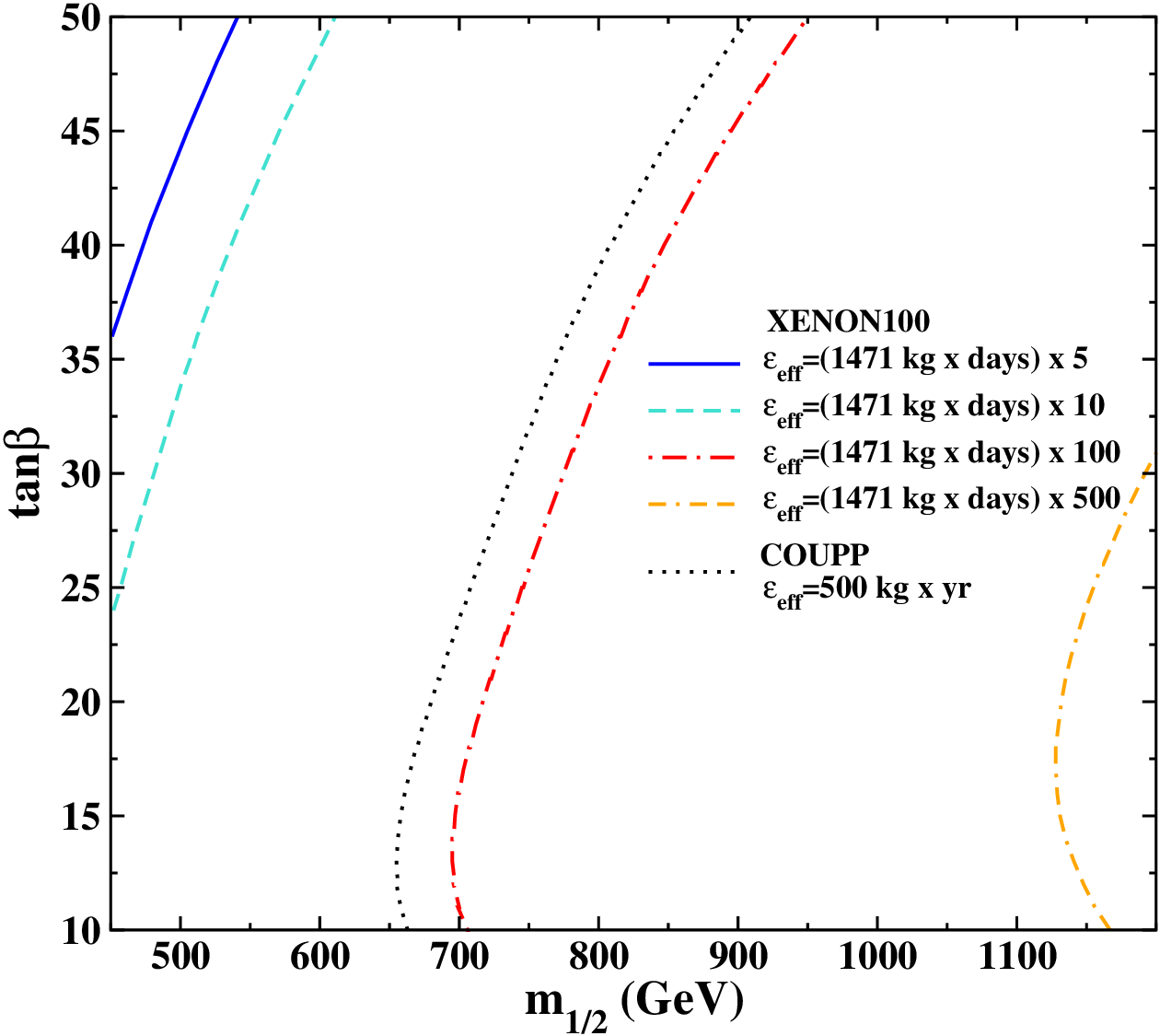}
\caption{Exclusion curves in the plane ($m_{1/2}, \tan\beta$) for the CMSSM
stau co-annihilation region with $A_0 =0$, $\mu>0$, set by an upper limit
on the spin-independent neutralino nucleon cross section using the fitting formulas 
Eq.~(\ref{fit}),~(\ref{massfit}). The regions to the left of the curves are excluded. 
The dotted line corresponds to the
upper limit of COUPP(500 kg$\times$yr), dashed blue line in Fig.~(\ref{fig4:COUPP})b.
The other lines are extrapolations for XENON100
where the present effective exposure is multiplied by factors 5, 10, 100, 500.}
\label{fig5:newlim}
\end{figure}

The result of this procedure is shown in Fig.~\ref{fig5:newlim}, where the excluded regions
are on the left of the curves. 
The COUPP upper limit with an effective exposure of 500 kg$\times$year, dashed blue line
in Fig.~(\ref{fig4:COUPP}a), corresponds to the dotted black line in Fig.~(\ref{fig5:newlim}).
The other curves are obtained for XENON100 considering an effective exposure to be 
5, 10, 100, 500 times the present value of 1471 kg$\times$days.
The dot-dashed red line, corresponds roughly to the effective exposure
of a future  ton mass detector with 1 year operation and total acceptance cut of 40\%.
The extrapolation of COUPP, dotted black line, is obtained without any acceptance cut.

We have to remark the limitations of the fitting formula. 
The coefficients in Table~\ref{table} have many particle 
physics uncertainties. First of all the cross section and the relic density
were calculated with \textsf{DarkSUSY} with the default values of the hadronic matrix elements.
Other codes can give slightly different values of the
cross section for the same input parameters. Furthermore the dependence
of SI cross section on not precisely known hadronic physics quantities can cause variations
up to a factor five for a given point of the CMSSM parameter space~\cite{ellisUncertain}.
There is a further  dependence of the SI cross section on $A_0$.
Anyway the choice $A_0 =0$ is the benchmark case study also for direct searches of 
supersymmetric particles at LHC: ATLAS and CMS typically present exclusion curves in the 
($m_0, m_{1/2}$)  plane with $A_0 =0$ and fixed $\tan\beta$~\cite{atlas,cms}.
With the proposed formula, hence, one has a direct idea  of the sensitivity
of a direct detection  experiment to one of the cosmologically favored region of the CMSSM parameter space
in a complementary way to LHC.

\section{Summary}
\label{sec6}

In this paper we have reviewed the formalisms and the approximations found in literature for the treatment 
of the SD neutralino-nucleus elastic scattering. We argued that 
all that one needs to correctly take into account 
the detailed nuclear physics information provided by shell-model calculations is just 
one of the normalized structure functions of Ref.~\cite{divari}.

We have shown that the factorization of the particle physics 
degrees of freedom from the nuclear physics momentum dependent structure functions
implied by this formalism
allows for a straightforward proof of the general formula ~(\ref{diseg}) proposed in Ref.~\cite{tovey}
without the need of the assumptions that were criticized in Refs.~\cite{BK}.

We have further discussed the ability of some of the present experiments and their future
upgrade to larger active masses (COUPP and XENON100) to constrain the stau co-annihilation
region of the CMSSM. In this region of the parameter space the neutralino mass is in the 
interval 180-550 GeV and the SI cross section is a decreasing function of the mass for $10<\tan\beta<50$,
taking values in the range $10^{-8}-10^{-10}$ pb and it is still poorly constrained by experiments. 
The SD cross sections, with the proton and the neutron, are in the 
range $10^{-6}-10^{-8}$ pb. 

COUPP,
although the high sensitivity of \fluorine ~to the proton SD scattering and the fact that 
the SD neutralino-nucleon cross sections are larger than the SI neutralino-nucleon cross section,  
can constrain the model in its large mass
phase only by the SI interaction with \iodine. The reasons are various:
first, because of the $A^2$ scaling of the SI neutralino-nucleus cross section; second,
there is a strong cancellation between the protons and neutrons contribution in 
the SD neutralino-\iodine ~cross section; third, the active mass of \fluorine ~is
small. 

Furthermore, in \fluorine , for the considered particle physics model, the SI rate 
is never negligible compared to the SD rate. In the case that the exposure were such
that the model could be probed through SD scattering, an exclusion curve in the 
plane ($m_\chi$, $\sigma^{\text{SD}}_p$) would be inaccurate.

Finally we have  given a fitting formula for the SI neutralino-nucleon cross section in the stau
co-annihilation region as 
a function of the two fundamental parameters $\tan\beta$ and $m_{1/2}$  ($10<\tan\beta<50$)
that allows to directly convert an upper limit
into an exclusion plot in the ($\tan\beta$, $m_{1/2}$) plane for the case study $A_0 =0$.

\acknowledgments

The author thanks J.~D.~Vergados and M.~E.~Gomez for inspiring discussions, suggestions and
comments on the manuscript. P.~Gondolo is acknowledged for useful questions during the TeVPA 2011 conference
where preliminary results of this paper were presented.
Support by MultiDark, grant CSD2009-00064 of the
Spanish MICINN project Consolider-Ingenio 2010 Programme and
partial support from MICINN project FPA2009-10773, MICINN-INFN(PG21)FPA2009-10773 and
from Junta de Andalucia under grant P07FQM02962 are acknowledged.

\appendix*

\section{Proof of the formulas of Section~\ref{sec2}}

In this Appendix we derive the formulas given in Section~\ref{sec2} following 
the formalism of Ref.~\cite{divari} but with a simplified and slightly 
different notation.

The neutralino-nucleon SD cross section
is determined by the axial part of effective lagrangian. At the nucleon level,
in the isospin representation that is convenient for nuclear physics calculations, we can write
\begin{equation}
\mathcal{L}_{eff}=\bar{\chi}\gamma^\mu \gamma^5 \chi \bar{N} 2s_\mu \frac{1}{2}(a_0\openone +a_1 \hat{\tau}_3) N.
\end{equation}
The  operator $\hat{\tau}_3$ act as $\hat{\tau}_3 |p\rangle =|p\rangle$, $\hat{\tau}_3 |n\rangle =-|n\rangle$ and $\openone$
is the identity operator in isospin space. Thus for $N=p,n$ the isospin operator gives $(a_0 \pm a_1)/2 =a_{p,n}=\sum_q d_q \Delta q^{(N)}$
where $d_q$ is the effective coupling with quarks and $\Delta q^{(N)} $ the spin fractions of the nucleon carried by the quarks.
We do not discuss further the physics involved at the nucleon level, see Refs.~\cite{ellisflores,ellisUncertain}.
 
Taking the non-relativistic limit we get the neutralino-nucleus spin-spin interaction  
\begin{equation}
\hat{V}=4 \hat{\mathbf{s}}_{\chi} \cdot \sum_{i=1}^{A}\frac{1}{2}(a_0\openone  +a_1 \hat{\tau}^3_i ) \hat{\mathbf{S}}_{i}
\delta(\mathbf{r}-\mathbf{r}_i).
\end{equation}
Here $\hat{\mathbf{S}}_{i}$ and $\mathbf{r}_i$ are the spin and coordinates of the $i$--th nucleon. In literature sometimes factors 
$G_F /\sqrt{2}$, $G_F \sqrt{2}$ or $G_F 2\sqrt{2}$ are extracted from $a_{0,1}$. To simplify the formulas, we adopt instead the convention 
that all the couplings are included in $a_{0,1}$.

The spin operator of the neutralino operates on eigenstates of the spin $|s\rangle$, while
conventionally all the angular momentum operators of the nucleus are evaluated in state with the maximal value of the $z$ projection,
$\langle \hat{\mathbf{O}}\rangle \equiv \langle J,M_J =J|\hat{O}^z|J,M_J =J \rangle$. 
The nuclear wave function depends also on the isospin and the coordinates
of the nucleons
\begin{equation}
|A\rangle =|J, M_J =J, \tau_3, \mathbf{r}_1...\mathbf{r}_A\rangle.
\end{equation}

The elastic differential cross section
in the center of mass frame and the total cross section, in the case that there is no angular dependence of the amplitude, are given by 
\begin{equation}
\frac{d\sigma}{d\varOmega}=\frac{\mu^2_A}{4\pi^2}\overline{|\mathcal{M}|^2},\,\,\,\,\sigma=\frac{\mu^2_A}{\pi}\overline{|\mathcal{M}|^2},
\label{dQM}
\end{equation}
where the scattering matrix element, with $|A, \chi\rangle =|A\rangle|s\rangle$, is 
\begin{equation}
\mathcal{M}=\langle A,\chi |\int d\mathbf{r} e^{-i\mathbf{q}\cdot \mathbf{r}} \hat{V}|A,\chi   \rangle.
\end{equation}
 
For two spin operators acting on different spaces, the average over the initial directions of modulus squared of the scalar product 
is $\overline{|\mathbf{S}_a \cdot \mathbf{S}_b|^2}=\frac{1}{3} \mathbf{S}^2_a \mathbf{S}^2_b$, hence
\begin{equation}
\overline{|\mathcal{M}|^2}=\frac{1}{3} 16 \langle \hat{\mathbf{s}}^2_{\chi}\rangle_s \langle \hat{\boldsymbol{\varSigma}}^2 \rangle_A,
\label{Mavaraged}
\end{equation}
We have defined the operator, 
\begin{equation}
\hat{\boldsymbol{\varSigma}}=\sum_{i=1}^{A}\frac{1}{2}(a_0\openone  +a_1 \hat{\tau}^3_i ) \hat{\mathbf{S}}_{i}
e^{-i\mathbf{q}\cdot \mathbf{r}_i} =\frac{1}{2}(a_0 \hat{\bm{\Omega}}_0 + a_1 \hat{\bm{\Omega}}_1),
\label{sigma}
\end{equation}
with
\begin{equation}
\hat{\bm{\Omega}}_0=\sum\limits_{i=1}^{A} \openone
\hat{\mathbf{S}}_i e^{-i\mathbf{q}\cdot \mathbf{r}_i},\;\;\;\;
\hat{\bm{\Omega}}_1=\sum\limits_{i=1}^{A} \hat{\tau}_i^3 
\hat{\mathbf{S}}_i e^{-i\mathbf{q}\cdot \mathbf{r}_i}.
\label{sigmadecomposed}
\end{equation}

To evaluate $\langle \hat{\boldsymbol{\varSigma}}^2 \rangle_A$ we note
that for a vector operator, the matrix elements in states $|J,M_J =J\rangle$ are 
related to the reduced matrix elements by~\cite{landau}
\begin{eqnarray}
\langle J||\hat{\mathbf{O}}||J \rangle&=&\frac{\sqrt{J(J+1)(2J+1)}}{J} \langle J,J|\hat{O}^z|J,J \rangle,\nonumber\\
\langle J,J|\hat{\mathbf{O}}^2|J,J \rangle &=& \frac{1}{2J+1}|\langle J||\hat{\mathbf{O}}||J \rangle|^2.
\label{wig2}
\end{eqnarray}
It follows:
\begin{equation}
\langle J,J|\hat{\mathbf{O}}^2|J,J \rangle =\frac{J+1}{J} |\langle J,J|\hat{O}^z|J,J \rangle|^2.
\label{wig3}
\end{equation}
We thus define the momentum dependent matrix elements
\begin{equation}
\Omega_0 (q)=2\sqrt{\frac{J+1}{J}}\langle 
\hat{\Omega}_0^z
\rangle_A ,\;\;
\Omega_1 (q)=2\sqrt{\frac{J+1}{J}}\langle 
\hat{\Omega}_1^z
\rangle_A .
\label{omega2q}
\end{equation}
From Eqs.~(\ref{sigma})--(\ref{omega2q}) we find 
\begin{equation}
\langle\hat{\boldsymbol{\varSigma}}^2\rangle_A=\frac{1}{16}|a_0 \Omega_0 (q)+a_1 \Omega_1 (q)|^2 .
\label{o2}
\end{equation}
Obviously,
$\langle \hat{\mathbf{s}}^2_{\chi}\rangle_s =s(s+1)={3}/{4}$.
Eq.~(\ref{Mavaraged}) thus takes the form
\begin{equation}
\overline{|\mathcal{M}|^2}=\frac{1}{4}|a_0 \Omega_0 (q)+a_1 \Omega_1 (q)|^2.
\label{M2verg}
\end{equation}
Expanding the square and factoring out the zero momentum values, we introduce the normalized structure functions 
$F_{ij}(q)$:%
\begin{equation}
F_{ij}(q)=\frac{\Omega_i(q) \Omega_j (q)}{\Omega_i(0) \Omega_j (0)}.
\label{Fdef}
\end{equation}
and find
\begin{eqnarray}
\overline{|\mathcal{M}|^2}
&=&\frac{1}{4}(a^2_0 \Omega^2_0 (0) F_{00}(q)+2 a_0 a_1 \Omega_0 (0) \Omega_1 (0) F_{01}(q)\nonumber\\
&+&a^2_1 \Omega^2_1 (0) F_{11}(q)).
\end{eqnarray}

By reason of Eq.~(\ref{Fequiv}), we can make the approximation
\begin{equation}
\overline{|\mathcal{M}|^2}\simeq\frac{1}{4}(a_0 \Omega_0 (0)+a_1 \Omega_1 (0))^2 F_{11}(q).
\label{a15}
\end{equation}
Taking $q=0$ in Eqs.~(\ref{omega2q}) and using $\hat{\tau}_3 |p\rangle =+|p\rangle$
and $\hat{\tau}_3 |n\rangle =-|n\rangle$ we have
\begin{eqnarray}
\Omega_0 (0)&=&2\sqrt{\frac{J+1}{J}}\langle \sum\limits_{i=1}^{A}\openone \hat{S}_i^z \rangle_A 
=2\sqrt{\frac{J+1}{J}}(\langle {\mathbf{S}}_p\rangle + \langle {\mathbf{S}}_n\rangle)\nonumber\\
&=&\Omega_p (0)+\Omega_n (0),\\
\Omega_1 (0)&=&2\sqrt{\frac{J+1}{J}}\langle \sum\limits_{i=1}^{A} \hat{\tau}^i_3 
\hat{S}_i^z  \rangle_A 
=2\sqrt{\frac{J+1}{J}}(\langle {\mathbf{S}}_p\rangle - \langle {\mathbf{S}}_n\rangle)\nonumber\\
&=&\Omega_p (0)-\Omega_n (0).
\end{eqnarray}
Eq.~(\ref{opn}) is thus proved.

Furthermore, using $a_{0,1} =a_p \pm a_n$ and Eq.~(\ref{opn}), Eq.~(\ref{a15}) becomes:
\begin{equation}
\overline{|\mathcal{M}|^2}= 4\frac{J+1}{J}(a_p \langle {\mathbf{S}}_p\rangle + a_n \langle {\mathbf{S}}_n\rangle)^2   F_{11}(q).
\label{M2standard}
\end{equation}
For a single nucleon (\ref{M2standard}) reduces to $3|a_{p,n}|^2$, hence 
\begin{equation}
\sigma^{SD}_{p,n}=3\frac{\mu^2_p}{\pi}|a_{p,n}|^2, 
\label{sigpn}
\end{equation}
Eq.~(\ref{sigSD}) follows from Eqs.~(\ref{M2standard})--(\ref{sigpn}).

Finally, with the substitution 
$d\varOmega = \frac{4\pi}{4\mu^2_A v^2} dq^2 =\frac{2m_A \pi}{\mu^2_A v^2} dE_R$
in Eq.~(\ref{dQM}) and using Eqs.~(\ref{M2standard}) and  (\ref{sigSD})
we obtain Eq.~(\ref{dsiggen}).

The present formalism and the standard formalisms are equivalent and connected by 
\begin{eqnarray}
S_{ij}(q)&=&\frac{2J+1}{(1+\delta_{ij})8\pi}\Omega_i(q) \Omega_j (q).
\label{Fequiv1}
\end{eqnarray}
For a given nuclear wave function
they furnish the same cross section. 

In last analysis, the difference resides
in the way by which the multipole decomposition  in vector spherical harmonics 
of the operator~(\ref{sigma}) is carried out. In the standard formalism this done 
in terms of the operator $\mathcal{T}^{el\,5}_L$ and $\mathcal{L}^{5}_L$, see Refs.~\cite{engel,engelrev,BS1}
and references therein for explicit formulas and meaning. 
Both the operators contain the couplings $a_0$ and $a_1$, thus  the modulus squared
of each contains terms proportional to $a^2_0$, $a^2_1$ and the interference $a_0 a_1$. 
The function $S(q)$ in terms of the functions $S_{ij}(q)$ arises after a rearrangement of these terms.

In the formalism of Ref.~\cite{divari} the multipole decomposition  in vector spherical harmonics is
done on the operators (\ref{sigmadecomposed}). Anyway, keeping separated the terms in $a_0$ and $a_1$ has the advantage that
the squared of the amplitude is always a perfect square, see
Eq.~(\ref{M2verg}), and limit $q=0$ is reached in a more transparent way.

\end{document}